\documentclass[
 reprint,
showpacs,
 amsmath,amssymb,
 aps,
prl,
]{revtex4-1}

\usepackage[all,matrix,arrow,curve]{xy}
\def\V{\mathbb{V}}
\def\W{\mathbb{W}}

\usepackage[pdftex]{graphicx}
\usepackage{dcolumn}
\usepackage{bm}

\def\be{\begin{equation}}
\def\ee{\end{equation}}
\def\ba{\begin{eqnarray}}
\def\ea{\end{eqnarray}}

\def\rd{\mathrm{d}}

\def\g{\mathfrak{g}}

\let\a=\alpha \let\b=\beta

\let\m=\mu

\let\e=\epsilon

\begin{document}

\title{ Leibniz-Yang-Mills Gauge Theories and the 2-Higgs Mechanism}

\author{Thomas Strobl}
\affiliation{Institut Camille Jordan,
Universit\'e Claude Bernard Lyon 1 \\
43 boulevard du 11 novembre 1918, 69622 Villeurbanne cedex,
France
}%
\begin{abstract}
 A quadratic Leibniz algebra $(\V,[ \cdot, \cdot ],\kappa)$ gives rise to a canonical Yang-Mills type functional $S$ over every space-time manifold. The gauge fields consist of 1-forms $A$ taking values in $\V$ and 2-forms $B$ with values in the subspace $\W \subset \V$ generated by the symmetric part of the bracket. If the Leibniz bracket is anti-symmetric, the quadratic Leibniz algebra reduces to a quadratic Lie algebra, $B\equiv 0$,  and $S$ becomes identical to the usual Yang-Mills action functional. We describe this gauge theory for a general quadratic Leibniz algebra. We then prove its (classical and quantum) equivalence to a Yang-Mills theory for the Lie algebra 
${\mathfrak{g}} = 
\V/\W$ to which one couples massive 2-form fields living in a ${\mathfrak{g}}$-representation. Since in the original formulation the $B$-fields have their own gauge symmetry, this equivalence can be used as an elegant mass-generating mechanism for 2-form gauge fields, thus providing a ``higher Higgs mechanism'' for those fields.

\end{abstract}

\maketitle

\section{Introduction}
Leibniz algebras are a simple generalization of Lie algebras, even though they have been much less studied than their anti-symmetric specialization. Recently \cite{talk,Waldram, Lavau,embedding} it became clear that the gauge sector within gauged supergravity theories constructed by means of the ``method of the embedding tensor'' \cite{H1,H2,H3,H4,H5,H6,H7} is based on precisely non-Lie Leibniz algebras. In such occurrences, however, one needs generically many more data and mathematical structures than the one of a Leibniz algebra \cite{HenningohneLie4,Sam,Lie4,gerbe,Wagemann}. 

It is thus remarkable, that already a Leibniz algebra alone can be used to define a consistent generalization of a Yang-Mills functional. Implicitly this is already a consequence of  \cite{gerbe}: the functional \eqref{S}  and its gauge transformations \eqref{symm1}, \eqref{symm2} below result from a specialization and simplification of formulas in \cite{gerbe} for the case of ordinary Leibniz algebras. In the present paper we highlight this particular case and work out details and consequences. 

First we present the gauge fields, their field strengths, and, last but not least, the infinitesimal gauge symmetries, using nothing else but the one given Leibniz relation, equation \eqref{Leibniz} below. Therefore, the corresponding calculations and framework can be useful also in the context of other, more general considerations, not relying on the particular functional we study here, namely the  Leibniz-Yang-Mills functional $S$. Its definition requires an additional non-degenerate bilinear form $\kappa$, for which gauge invariance induces a suitable compatibility condition, namely equation \eqref{kappa} below. This additional structure will be shown to restrict the original Leibniz algebra considerably, requiring a Loday cohomology class, characterizing the given Leibniz algebra otherwise, to be zero. As we will see, this in turn makes it possible to reformulate the theory in purely Lie algebraic terms. What we find is that in this reformulation, one obtains an ordinary Yang-Mills gauge theory for a Lie algebra $\g$, to which, however, 2-form fields are coupled in a $\g$-representation. Somewhat surprisingly, we find that they now have an induced mass, all of them. This is physically interesting for several reasons, one of which being that if the 2-form fields were massless, we should have already observed them in experiments.

The gauge  theory presented here may be viewed upon also as a particular higher gauge theory \cite{Baez}. It can therefore be related to $L_\infty$-algebras \cite{Lada1,Lada2} for the description of  which super-geometrical methods are most suitable  \cite{Gruetzmann,Qbundles}. We explicitly avoid this language here so as to keep the presentation  self-contained and  more widely accessible. We remark at this point, however, that while the presence of an $L_\infty$-algebra is  necessary  for the construction of a higher gauge theory \cite{Gruetzmann}, it is not sufficient, at least if one wants  non-trivial interactions between different form-degree gauge fields. Correspondingly, although every Leibniz algebra is shown to give rise to an $L_\infty$-algebra in a canonical way in \cite{embedding}, this does not yet guarantee the existence of a consistent and non-trivial gauge invariant functional for gauge fields of arbitrarily 
high form degree. The present construction is only the first step into the direction of the conjecturally present hierarchy of higher gauge theories given for every Leibniz algebra with adequate bilinear forms. 

The structure of the paper is as follows: We first present the gauge fields (``connections''), their field strengths (``curvatures''), and the Leibniz-Yang-Mills functional. The gauge symmetries and a proof of gauge invariance of $S$ is subject of the subsequent section. The techniques and most formulas presented there hold true for every Leibniz algebra; only in the very last step, the invariance of $S$, the existence of a compatible $\kappa$ is used. In the ensuing section, however, we make use precisely of the restrictions posed on a Leibniz algebra by $\kappa$ and turn to the main finding of this paper, 
namely the above-mentioned equivalence of the Leibniz-Yang-Mills functional with a Yang-Mills theory coupled to massive 2-form fields that one obtains by an appropriate field redefinition. After some reflections of a more mathematical nature,  we conclude with a  summary and outlook.

\section{Leibniz algebras, the gauge fields, and the functional}
Let $(\V,[ \cdot, \cdot ])$ be a Leibniz algebra, i.e.~a vector space $\V$ equipped with a bilinear product, denoted by a bracket, which satisfies
\be [x,[y,z]] = [[x,y],z] + [y,[x,z]] \label{Leibniz}
\ee 
for all $x,y,z \in \V$. We call it quadratic, if it carries a non-degenerate symmetric bilinear form $\kappa$ satisfying the obvious invariance condition
\be \kappa([x,y],z) + \kappa(y,[x,z]) = 0. \label{kappa}
\ee 
If the bracket is anti-symmetric, $[x,y]=-[y,x]$, a (quadratic) Leibniz algebra is a (quadratic) Lie algebra. 

The Yang-Mills functional defined for every quadratic Lie algebra plays an essential role in the high energy physics of elementary particles. Here we show that this functional can be defined already for a quadratic Leibniz algebra, without the need to specify any additional data except for the choice of a spacetime $(M,g)$. Its definition reduces to the usual Yang-Mills theory precisely for those Leibniz algebras which are Lie algebras. 

For its description, we first split the bracket into its symmetric and anti-symmetric parts,
\be [x,y]_\pm := \tfrac{1}{2} \left([x,y] \pm [y,x] \right).
\ee
Let us denote the vector space generated by the image of the symmetric part by $\W$, 
\be \W := \{x\in \V \vert \exists y_i,z_i \in \V, x =\sum_i [y_i,z_i]_+\} \, .\label{W}
\ee
The gauge fields of the Leibniz-Yang-Mills theory are
\be A \in \Omega^1(M,\V)\qquad \mathrm{and} \qquad B \in \Omega^2(M,\W) \: , \ee
i.e.~1-forms and 2-forms on $M$ with values in $\V$ and $\W$, respectively. Evidently, if the symmetric part of the bracket vanishes, and only then, the Leibniz algebra becomes a Lie algebra, $\W$ becomes the 0-vector inside $\V$, and these gauge fields reduce to the Lie algebra valued connection 1-forms we are used to from ordinary Yang-Mills gauge theories. Next we define the generalization of the curvature 2-forms $F$:
\ba F &=& \rd A + \tfrac{1}{2}[A,A] - B \,,\label{F}\\
G& =& \rd B + [A, B-F+ \tfrac{1}{6} [A,A]]_+ \, .\label{G}
\ea 
To avoid any misunderstanding, some remarks on the notation: If there is a product of differential forms, a wedge product is understood. Thus, using a basis $e_a$ of $\V$, we have, e.g., $A=A^a \otimes e_a$ for (ordinary) 1-forms $A^a=A^a_\m(x) \rd x^\m$, and a term like $[A,A]$ stands for $A^a \wedge A^b \otimes [e_a,e_b]$. 
Since $\W\subset \V$, subtracting $B$ from the other terms in \eqref{F} is a meaningful operation. We observe that for $\V = \mathfrak{g}$, a Lie algebra, not only $B=0$, but also $G=0$, since the symmetric part of the bracket vanishes identically in this case. The above definitions imply 
\be F \in \Omega^2(M,\V)\qquad \mathrm{and}  \qquad G\in \Omega^3(M,\W)\, .\ee 
Now we are ready to present the Leibniz-Yang-Mills action functional:
\ba S[A,B] &=& \tfrac{1}{2} \int_M \left( \parallel \!F \!\parallel^2 + \parallel \!G \!\parallel^2 \right) \label{S}\\ 
&\equiv & \tfrac{1}{2} \int_M \kappa(F ,\ast F)+ \kappa(G,\ast G)\, . \nonumber \ea
Here $\ast$ denotes the usual Hodge star operation induced by the pseudo-Riemannian metric $g$ on $M$; e.g.,  \be \kappa(F,\ast F) =  \kappa_{ab} F^a \wedge \ast F^b =  F_{\mu \nu}^a F^{b,\mu\nu} \kappa_{ab} \, \sqrt{\vert \det g \vert}\rd^nx \, . \ee 
Note that we can use the same pairing $\kappa$ for $F$ and $G$, provided only that its restriction to $\W$ is non-degenerate.  Under such an assumption we call $(\V,[\cdot,\cdot],\kappa)$ a \emph{split-quadratic Leibniz algebra}. In the physically most relevant case of a positive definite $\kappa$, this is certainly always satisfied, and we assume it otherwise.  If one wishes so, one may want to also introduce two coupling constants $\lambda_F$ and $\lambda_G$, multiplying the two contributions to $S$ appropriately, tuning their relative weight and possibly introducing the appropriate physical dimensions.

\section{\label{sec:gauge} Gauge symmetry and invariance} 

The infinitesimal gauge transformations of the fields $A$ and $B$ are parametrized by $\V$-valued functions $\epsilon$  on $M$ and $\W$-valued 1-forms $\mu$. They are of the form
\ba \delta A &=& \rd \epsilon + [A, \epsilon]_- + \mu \, ,
\label{symm1} \\ 
\delta B &=& \rd \mu + [A ,\mu]_+ + [F-B,\epsilon]_+ -\tfrac{1}{2}[A ,[A ,\epsilon]]_+\, .\label{symm2}
\ea  
We will now show that they induce the following transformations of \eqref{F} and 
\eqref{G}: 
\be \delta F = -[\epsilon,F] \quad , \qquad    \delta G = -2 [\epsilon,G]_+\, . 
\label{deltaF}\ee 
Before we start the calculations, we specify some of the rules that are valid in the concise notation used here: As indicated  before, for every two $\V$-valued differential forms $\alpha$ and $\beta$, an expression like $[\alpha,\beta]_\pm$ stands, more explicitly, for $\alpha^a \wedge \beta^b \otimes [e_a,e_b]_\pm$. This has, e.g., the following consequence:
\be  [\alpha,\beta]_\pm = \tfrac{1}{2} \left(  [\alpha,\beta] \pm (-1)^{|\alpha|\, |\beta|}[\beta,\alpha]\right) \, , 
\ee
where $|\alpha|$ and $|\beta|$ denote the form-degrees of $\alpha$ and $\beta$, respectively. In particular, if both forms are  odd:  $[\alpha,\beta]_\pm = \tfrac{1}{2} \left(  [\alpha,\beta] 
\mp [\beta,\alpha]\right)$. In addition, one verifies
\ba  [\alpha,\beta]_\pm &=& \pm (-1)^{|\alpha|\, |\beta|}[\beta,\alpha]_\pm \, , \\
\rd  [\alpha,\beta] &=&  [\rd \alpha,\beta] + (-1)^{|\alpha|}[\alpha,\rd \beta] \, ,\\
\rd  [\alpha,\beta]_\pm &=&  [\rd \alpha,\beta]_\pm + (-1)^{|\alpha|}[\alpha,\rd \beta]_\pm \, , \label{dpm}
\ea 
while the defining Leibniz condition \eqref{Leibniz} turns into:
\be [\a,[\b,\gamma]]-(-1)^{|\alpha|\, |\beta|} [\b,[\a,\gamma]]=[[\a,\b],\gamma]\, .\label{abg}\ee 
This shows in particular that $(\Omega^\bullet(M,\V),\rd, [\cdot, \cdot ])$ is a \emph{graded differential Leibniz algebra}, with $ [\cdot, \cdot ]_+$ and  $ [\cdot, \cdot ]_-$ the \emph{graded} symmetric and anti-symmetric parts of the bracket. These two parts of the bracket do not satisfy the relation \eqref{abg}; instead, one has, for example,
\ba  [\a,[\b,\gamma]_-]_--[[\a,\b]_-,\gamma]_-&-&(-1)^{|\alpha|\, |\beta|} [\b,[\a,\gamma]_-]_- \nonumber\\ &=&-\mathrm{Alt}[\a,[\b,\gamma]]_+
  \, , \label{Lie2} \ea
where 
$\mathrm{Alt}$ denotes the graded anti-symmetrization projector. 
In fact, with \cite{embedding} one shows that the 2-term complex $\Omega^\bullet(M,\W) \to \Omega^\bullet(M,\V)$ can be endowed with the structure of a differential graded Lie 2-algebra whose non-vanishing 2-bracket is given by $ [\cdot, \cdot ]_-$; but since we want to avoid such terminologies in the present paper for simplicity of the presentation, we will not go further into this perspective here.

We now decompose the gauge transformations into two parts, $\delta = \delta_\epsilon + \delta_\mu$, and first consider (the degree zero derivation) $\delta_\mu$, where $\delta_\mu A = \mu$ and $\delta_\mu B =  \rd \mu + [A ,\mu]_+$. This yields
\be  \delta_\mu F =  \rd \mu + \tfrac{1}{2}[A,\mu] +  \tfrac{1}{2}[\mu ,A ] -  \rd \mu - 
[A ,\mu]_+ = [\mu ,A ]\, . \ee
Now it is important that $\mu$ takes values in $\W$, $\mu \in \Omega^1(M,\W)$, and that this subspace lies in the left-center of the Leibniz algebra, $\W \subset Z_L(\V)$: Adding to \eqref{Leibniz} the same equation with $x$ and $y$ exchanged, we see that for every $w = [x,y] + [y,x] \in \W$ one has $[w,z]=0$ for all $z\in \V$. This then proves $\delta_\mu F=0$. 

Henceforth we will freely use $[\mu, \cdot ]=0= [B,\cdot]$, such that, e.g., $\delta_\mu B =  \rd \mu + \tfrac{1}{2}[A ,\mu]$. This also implies that the second equation in \eqref{deltaF}, adapted to show explicitly that the change of $G$ lies inside $\W$, can be rewritten also as
\be \label{deltaG} \delta G =-[\epsilon,G]  \, . 
\ee
Similarly, in some calculations it is useful to note that $[[\alpha,\beta]_+,\cdot ]=0$, which implies, e.g., \be [[\alpha,\beta],\gamma]=-(-1)^{|\alpha|\, |\beta|}[[\beta,\a],\gamma]\, . \ee

We now turn to showing $\delta_\mu G=0$. Straightforward and simple calculations yield 
\ba \delta_\mu (\rd B ) &=& \tfrac{1}{2} \left( [F-\tfrac{1}{2} [A,A],\mu] - [A,\rd \mu ]\right),  \\
\delta_\mu ([A, B-F]_+) &=& \tfrac{1}{2} \left( [A,\rd \mu + \tfrac{1}{2}[A,\mu]]-[F,\mu] \right),  \\
\delta_\mu ([A,\tfrac{1}{6} [A,A]]_+)&=& \\ 
&& \hspace{-15mm}=\tfrac{1}{12} \left([[A,A],\mu]+[A,[A,\mu]]+[[A,\mu],A] \right).\nonumber 
\ea 
Taking the sum of these three lines, we obtain
\be \delta_\mu G = -\tfrac{1}{6}[[A,A],\mu]+\tfrac{1}{3}[A,[A,\mu]]+\tfrac{1}{12}[[A,\mu],A]\, . \ee
The first two terms on the r.h.s.\ of this equation cancel due to 
$[[A,A],\mu]=2[A,[A,\mu]]$, which follows from \eqref{abg}. And the last term vanishes on the nose, since $\W$ is an ideal of $\V$ (in particular $[\V,\W]\subset \W$, and thus $[A,\mu]\in\Omega^2(M,\W)$), as one shows by symmetrizing \eqref{Leibniz} with respect to $y$ and $z$:
\be [x,[y,z]_+]=[[x,y],z]_+ + [y,[x,z]]_+ \, . \label{ideal}
\ee  

The transformation behavior of $F$ and $G$ with respect to $\delta_\epsilon$ can now be proven in a very similar fashion by these techniques. We restrict ourselves to showing only the first equation in \eqref{deltaF} in detail, while leaving \eqref{deltaG} as an exercise to the reader. We start as follows
\be \delta_\epsilon \left(\rd A + \tfrac{1}{2}[A,A] \right)= [\rd A,\epsilon]_-+[A,[A,\epsilon]_-]_-\, . \ee
Now eliminate on both sides $\rd A$ by means of \eqref{F}; thus, e.g., the l.h.s.\ becomes $\delta(F+B)$. We are showing this calculation also since it permits to in fact \emph{deduce} the $\epsilon$-part of \eqref{symm2}, provided the first equation in  \eqref{deltaF} holds true: Using $[F+B,\epsilon]_- = - [\epsilon,F] + [\epsilon,F-B]_+$, one finds the necessary change of $B$ to be of the given form, provided only
\be -\tfrac{1}{2}[[A,A],\epsilon]_-+[A,[A,\epsilon]_-]_-= -\tfrac{1}{2}[A ,[A ,\epsilon]]_+\, ,\ee 
which one verifies by using \eqref{Lie2}. 

It remains to show the gauge invariance of \eqref{S}. The condition \eqref{kappa} yields
\be  \kappa([\alpha,\beta],\gamma) + (-1)^{|\alpha|\, |\beta|}  \kappa(\beta,[\alpha,\gamma])=0 \, , \ee
from which we deduce
$  \kappa([\epsilon,\beta],\ast\beta) +  \kappa(\beta,[\epsilon,\ast \beta])=0 $ for every $\epsilon\in C^\infty(M,\V)$, $\beta \in \Omega(M,\V)$. Together with \eqref{deltaF} and \eqref{deltaG}, this then indeed proves 
\be \delta S = 0 \, . \ee

\section{Equivalence to standard Yang-Mills with massive 2-form fields\label{equivalence}}
It is time to study the notion of a split-quadratic Leibniz algebra into further depth. Recall, that in addition to the condition \eqref{kappa}, which is needed for gauge invariance, we require the restriction $\kappa|_\W$ of $\kappa$ to $\W$ to be non-degenerate, which is needed for the non-trivial propagation of the $B$-gauge fields in \eqref{S}. This implies
\be \V = \W^\perp \oplus \W \: \ni (\xi,w) \, , \label{decomp}
\ee 
so that every element $v \in \W$ can be decomposed uniquely into a part in $\W^\perp$ and another one  in $\W$. For every $\xi,\xi' \in  \W^\perp$ and $w \in \W$, we have 
\be \kappa([\xi,\xi'], w) = - \kappa(\xi',[\xi,w])= 0, \ee 
since $[\V,\W]\subset \W$. Thus $\W^\perp$ is a subalgebra of $\V$. Consequently, since $[\xi,\xi']_+$ lies in $\W$ by definition, it has to vanish. The restriction of the Leibniz algebra to $\W^\perp$ is a \emph{Lie algebra}, which we denote by $(\g,[\cdot, \cdot]_\g)$ henceforth. Last but not least, the Leibniz property \eqref{Leibniz} yields in particular $[\xi,[\xi', w]] - [\xi',[\xi, w]] =  
[[\xi,\xi'], w]$, which proves that $\W$ is a $\g$-representation. Henceforth, we denote the $\g$-action on $\W$ given by the bracket from left, $[\xi,w]$, by the more conventional $\xi \cdot w$. Since, in addition, $[w,\xi]=0=[w,w']$, we have 
\be  [(\xi,w),(\xi',w')] = ([\xi,\xi']_\g,\xi \cdot w') \, . \label{noalpha}
\ee 
The decomposition \eqref{decomp} also applies to the bilinear form, $\kappa((\xi,w),(\xi',w'))= \kappa(\xi,\xi') + \kappa(w,w')$, and then the condition \eqref{kappa} shows that $\kappa$ is in one-to-one correspondence with a non-degenerate ad-invariant symmetric bilinear form on $\g$ and a likewise $\g$-invariant one on $\W$. Thus we find, in a slight generalization of a result established  in \cite{Wagemann}, that every split-quadratic Leibniz algebra is the same as an ordinary quadratic Lie algebra together with a quadratic representation:
\be (\V,[\cdot, \cdot],\kappa) = (\g,[\cdot, \cdot]_\g,\kappa_\g) \ltimes (\W, \kappa_\W) \, .
\ee 

We now want to show what consequences this has for the functional \eqref{S}. Let us for this purpose split the 1-form gauge field into its two parts, $A=A_\g + A_\W$. Then also $F=F_\g + F_\W$, where $F_\g = \rd A_\g + \tfrac{1}{2} [A_\g,A_\g]_\g$ is the ordinary curvature 2-form of the $\g$-connection $A_\g$ and, on the other hand, $F_\W = \rd A_\W + \tfrac{1}{2} A_\g \cdot A_\W - B$. Next we rewrite $G$ as 
\be  G = \rd B + [A,B] - [A,\rd A]_+-\tfrac{1}{2}[A,[A,A]] \, , \ee
where we made use of $[A,[A,A]]_+=\tfrac{3}{2}[A,[A,A]]$ and implement  the above decomposition of $A$:
 \be G = \rd B + A_\g \cdot B - \tfrac{1}{2} \rd A_\g \cdot A_\W- \tfrac{1}{2} A_\g \cdot \rd A_\W- \tfrac{1}{2} A_\g \cdot (A_\g \cdot A_\W) \, .\ee
Now introduce a new 2-form field by setting $\widetilde{B} := B - \rd A_\W - \tfrac{1}{2} A_\g \cdot A_\W$; since this change of fields has Jacobian one, it is also a valid operation at  the quantum level. A simple calculation then shows that in these new coordinates on field space, \eqref{S} takes the form
\ba S[A_\g,\widetilde{B}]&= & \int_M \Big( \kappa_\g(F_\g, \ast F_\g) \Big. \nonumber \\ 
&& \left. + \kappa_\W(D_{ \g} \widetilde{B},\ast D_{\g} \widetilde{B}) + \kappa_\W( \widetilde{B}, \ast \widetilde{B})\right)  \, ,\label{newS}
\ea
where $D_{ \g} \widetilde{B}:= \rd \widetilde{B} + A_\g \cdot \widetilde{B}$ is the \emph{standard} covariant derivative for a field taking values in a $\g$-representation $\W$. Most noteworthy, the action functional now does no more depend on $A_\W$, the part of the 1-form gauge field taking values in $\W \subset \V$. This independence can be identified with the $\delta_\m$-part of the gauge symmetries \eqref{symm1} and \eqref{symm2}. One may still wonder what happened to the $\W$-part of the $\delta_\e$-transformations. In fact, there is a subtlety in the previous parametrization of the gauge symmetries: For every choice of $\bar{\e}\in \Omega^0(M,\W)$, one has 
\be \delta_{\e:=\bar{\e}} + \delta_{\m:=\rd \bar{\e} + [A,\bar{\e}]_-} \equiv 0 \, . \label{reducible}
\ee 
In absence of a splitting \eqref{decomp} which also respects the bracket structure, it is useful to work with the reducible gauge symmetries \eqref{symm1} and \eqref{symm2}. Now, however, we can split the parameter $\epsilon$ into its two parts $\e_\g$ and $\e_\W$ and forget about the latter parameters, as their effect can be captured by means of a particular, field-dependent $\m$-shift. (Note that this is a legitimate  operation due to $ [A,\bar{\e}]_- = \tfrac{1}{2} A_\g \cdot \bar{\e} \in \Omega^1(M,\W)$). One verifies also that for $\e=\e_\g\in C^\infty(M,\g)$ and $\m=0$ the transformations  \eqref{symm1} and \eqref{symm2} induce 
the obvious symmetry of  \eqref{newS}:
\be \delta_{\e_\g} A_\g = \rd \e_\g + [A_\g,\e_\g] \, , \quad \delta_{\e_\g} \widetilde{B}= - \e_\g \cdot \widetilde{B} \, . \label{newsymm}
\ee 

Thus, at the classical level as well as at the quantum level, the Leibniz Yang-Mills gauge theory naturally associated to a split-quadratic Leibniz algebra turns out to be \emph{equivalent} to an ordinary Yang-Mills gauge theory with gauge fields $A_\g$ to which one couples massive 2-form fields $\widetilde{B}$ taking values in the $\g$-representation $\W$. The originally present $\delta_\m$-symmetry for the $B$-gauge fields disappeared now together with $A_\W$.

\section{\label{sec:new} Remarks on  finite gauge transformations} 

It is well known that the existence of an adinvariant, non-degenerate bilinear form poses restrictions on the chosen Lie algebra. A Lie algbra $\g$ permitting such a product $\kappa$ is called ``quadratic'' or ``metric''. We now want to illustrate that for a Leibniz algebra $\V$ the restriction to be ``(split-)quadratic'' or ``(split-)metric'' is even more drastic in some way.

 There is a short exact 
sequence of Leibniz algebras for every Leibniz algebra $\V$:
\be 0 \to \W \to \V \to \g  \to 0\, .\label{sequence} \ee
Here $\W$ is again defined as in \eqref{W} and $\g$ is the Lie algebra that one obtains as the quotient of $\V$ by its squares. Now, choosing an arbitrary splitting $\sigma \colon \g \to \V$ of the above sequence, we again can represent elements of $\V$ as couples $(\xi,w) \subset \g \otimes \W$, but now instead of equation \eqref{noalpha} one has 
\be  [(\xi,w),(\xi',w')] = ([\xi,\xi']_\g,\xi \cdot w' + \alpha(x,x')) \, .\label{alpha} \ee 
Here $\alpha \colon \g \otimes \g \to \W$ is a cocycle with respect to the Loday differential \cite{Loday1,Loday2}. Changing the splitting, leads to a modification of $\alpha$ by a coboundary of this differential. Thus, Leibniz algebras, in the picture of the above sequence, are characterized by a cohomology class $[\alpha]\in H^2_{Loday}(\g,\W)$ in general. 

As we showed in the previous section, the presence of a split-quadratic form $\kappa$ enforces this cohomology class $[\alpha]$ to vanish. It was this simplification, induced by the compatible $\kappa$, that made it possible to take complete recourse to Lie algebra techniques in the previous section.

At this point it is worth returning to the gauge symmetries \eqref{symm1} and \eqref{symm2}. They lead to the transformation properties 
\eqref{deltaF} for arbitrary Leibniz algebras. On the other hand, they were provided in infinitesimal form only. 

Addressing their integration to finite gauge transformations is a highly non-trivial question in the general setting. It evidently relates to the integration of Leibniz algebras \cite{W1,W2,W3}, and, simultaneously, of semi-strict Lie 2-algebras \cite{Lie21,Lie22,Lie23}  \footnote{Every Leibniz algebra gives rise to a Lie 2-algebra in a canonical way \cite{talk,LeibnizLie2,embedding}. A Lie 2-algebra is an 
$L_\infty$-algebra concentrated only in degrees 0 and -1. If it is strict, it corresponds to a crossed module of Lie algebras \cite{Baez2}, which in turn corresponds to a crossed module of Lie groups. The present Lie 2-algebra is, however, non-strict and there is no evident relation to groups then.}.  Both of these topics are subject to ongoing research in mathematics. One of the questions in this context is about what one wants or expects from the integration of these mathematical structures. The guiding principle of the relation of a Lie algebra to a Lie group is not completely sufficient for this. There can be different generalizations of the integration of a Leibniz algebra to some ``group-like manifold'' which all reduce to a Lie group for  the case that $\W=0$. Even worse, there can be equivalent notions to the one of a group---like a particular cocommutative Hopf algebra---which might lead to a completely different integrating global object for a Leibniz or Lie 2-algebra in the end. 

On the other hand, the integrated version of the gauge symmetries \eqref{newsymm} is standard. Thus, in the case of a split-quadratic Leibniz algebra, one should be able to trace the finite gauge transformations of $A_\g$ and $\widetilde{B}$ back to finite gauge transformations of the original fields $A$ and $B$. In this process, one has to reintroduce also the fields $A_\W$ together with their shift symmetry parametrized by $\mu$. And, if one wants to relate this to the infinitesimal transformations provided above, namely equations \eqref{symm1} and \eqref{symm2}, it will be necessary to reimplement also an integrated version of the redundancy condition \eqref{reducible}. 

While we will not comment on the finite gauge transformations here any further, we still hasten to add that it might be interesting to see the interplay of this question with the one of an integration of Leibniz algebras and semi-strict Lie 2-algebras. Moreover, the latter integration plays an important role for the construction of non-trivial bundles. And even in the context of \cite{Qbundles}, where the fiber of the bundle is the Lie 2-algebra, an integration of the infinitesimal symmetries is needed.

\section{Summary and Outlook \label{outlook}}

In this paper we highlighted the fact that a Yang-Mills functional does not only exist for every quadratic Lie algebra, but already for a split-quadratic Leibniz algebra. We presented an index-free formalism that can be used also for more general gauge theories related to Leibniz algebras. Our main result was to show that the gauge theory \eqref{S} can be recast into the form of a Yang-Mills-theory for a Lie algebra to which 2-form fields are coupled. In the reformulation, where the specific form of split-quadratic Leibniz algebras was used, cf.\ equations \eqref{noalpha} and \eqref{alpha}, the 2-form gauge fields turned into massive 2-form fields without an independent gauge invariance. We thus obtain a higher version of a Higgs mechanism: The gauge invariance \eqref{symm2} excludes the addition of explicit mass terms for the 2-form fields $B$, similarly to what we know for the usual 1-form gauge fields. But such as an appropriate gauge-invariant addition of 0-form fields can be used to effectively generate masses for the 1-form fields, here it is the 1-forms that generate them for the 2-form fields. Interestingly, this does not need to be implemented by an intricate coupling procedure, it is achieved \emph{for free} by considering the functional \eqref{S} based on a Leibniz 
algebra \footnote{We propose the name 2-Higgs mechanism for this. First, it is a higher version of an ordinary Higgs mechanism, and in the language of `higher structures'' one conventionally labels  a hierarchy of structures by an integer $n \in \mathbb{N}$, with $n=1$ denoting the original notion. Second, and more importantly, it gives masses to 2-form gauge fields and, more generally, an $n$-Higgs mechanism would do so for $n$-form gauge fields.}.

The presented equivalence was established under the assumption $\kappa$ is split-quadratic, i.e.\ non-degenerate on $\V$ and also upon restriction to $\W \subset \V$. One may want to see what happens when this requirement is relaxed. The presence of the $\m$-shift symmetry seems to suggest, however, that also for only quadratic Leibniz algebras such a relation to a Lie algebra Yang-Mills theory holds---at the quantum level upon an appropriate partial gauge fixing forcing the part of $A$ with values in $\W$ to vanish. Another, more drastic alternative might be to drop the non-degeneracy condition of $\kappa$. Consider, e.g., 
$\kappa|_\W \equiv 0$ \cite{LeibnizCS}: here $B$ drops out from the action, 
$S[A,B]=S[A]$. But then, and only then, also $S[A]=S[A+A_\W]$ for every $A_\W \in \Omega^1(M,\W)$. This in turn implies that 
 effectively $A$ takes values in the quotient $\V/\W=\g$, cf.\ the sequence \eqref{sequence}. Although again we can no longer split $A$ into two parts in the same way as before, we end up with the standard Yang-Mills gauge theory for the (quadratic) Lie algebra $\g$, but here without  2-form fields \footnote{In fact, such an argument should also apply to the Leibniz-Chern-Simons theory considered in \cite{LeibnizCS}.}. In the physically most relevant case of a positive definite $\kappa$, on the other hand, our assumptions are all satisfied.

 In a final paragraph we come back to the language of $L_\infty$-algebras, which we tried to touch upon as little as possible in this article for the reasons explained in the Introduction. We saw that there is a functional $S$ associated canonically to every quadratic Leibniz algebra. Underlying to it is a Lie 2-algebra, which results upon truncation of a Lie $\infty$-algebra (an $L_\infty$-algebra concentrated in non-positive degrees) associated to every Leibniz algebra \cite{embedding,talk}. It is 
suggestive that there exists a higher Yang-Mills-type functional for arbitrarily high form-degrees associated  to every Leibniz algebra, if the latter algebra is equipped  with an appropriately non-degenerate invariant bilinear form. Moreover, this functional should permit consistent truncations to arbitrary levels (highest form degrees). The functional $S$ presented here corresponds to  level 2,  level 1 is the Yang-Mills theory for the Lie algebra $\g$ without 2-form fields \footnote{As already evident from these examples at lowest levels, the truncation is not consistent, if one merely puts higher form-degree gauge fields to zero. Truncations always involve appropriate quotient constructions at the highest non-vanishing level. For example, the functional $S[A,0]$, 
resulting from \eqref{S} by setting the 2-form fields to zero, is not gauge invariant anymore. But taking the quotient of $\V$ by $\W$, cf.\ \eqref{sequence}, we end up with the usual Yang-Mills functional at level one, $S_{YM}[A_\g]$.}. One may expect that some of the higher gauge fields of such a theory, although probably not all of them, receive masses by a generalization of the mechanism found in this article. To be seen.

It would be also interesting to see, if and how the present analysis extends to the context of \emph{enhanced} Leibniz algebras \cite{gerbe,Wagemann}.

\acknowledgments
I am grateful to Alexei Kotov for discussions and for stimulation to write this article. I also want to thank the anonymous referee for a very careful reading of the original manuscript and for the suggestion to add remarks on finite gauge transformations. This work was supported by the LABEX MILYON (ANR-10-LABX-0070) of Universit\'e de Lyon.


\begin{thebibliography}{4}%
\makeatletter
\providecommand \@ifxundefined [1]{%
 \@ifx{#1\undefined}
}%
\providecommand \@ifnum [1]{%
 \ifnum #1\expandafter \@firstoftwo
 \else \expandafter \@secondoftwo
 \fi
}%
\providecommand \@ifx [1]{%
 \ifx #1\expandafter \@firstoftwo
 \else \expandafter \@secondoftwo
 \fi
}%
\providecommand \natexlab [1]{#1}%
\providecommand \enquote  [1]{``#1''}%
\providecommand \bibnamefont  [1]{#1}%
\providecommand \bibfnamefont [1]{#1}%
\providecommand \citenamefont [1]{#1}%
\providecommand \href@noop [0]{\@secondoftwo}%
\providecommand \href [0]{\begingroup \@sanitize@url \@href}%
\providecommand \@href[1]{\@@startlink{#1}\@@href}%
\providecommand \@@href[1]{\endgroup#1\@@endlink}%
\providecommand \@sanitize@url [0]{\catcode `\\12\catcode `\$12\catcode
  `\&12\catcode `\#12\catcode `\^12\catcode `\_12\catcode `\%12\relax}%
\providecommand \@@startlink[1]{}%
\providecommand \@@endlink[0]{}%
\providecommand \url  [0]{\begingroup\@sanitize@url \@url }%
\providecommand \@url [1]{\endgroup\@href {#1}{\urlprefix }}%
\providecommand \urlprefix  [0]{URL }%
\providecommand \Eprint [0]{\href }%
\providecommand \doibase [0]{http://dx.doi.org/}%
\providecommand \selectlanguage [0]{\@gobble}%
\providecommand \bibinfo  [0]{\@secondoftwo}%
\providecommand \bibfield  [0]{\@secondoftwo}%
\providecommand \translation [1]{[#1]}%
\providecommand \BibitemOpen [0]{}%
\providecommand \bibitemStop [0]{}%
\providecommand \bibitemNoStop [0]{.\EOS\space}%
\providecommand \EOS [0]{\spacefactor3000\relax}%
\providecommand \BibitemShut  [1]{\csname bibitem#1\endcsname}%
\let\auto@bib@innerbib\@empty
\bibitem [{Note1()}]{Note1}%
  \BibitemOpen
  \bibinfo {note} {Every Leibniz algebra gives rise to a Lie 2-algebra in a
  canonical way \cite {talk,LeibnizLie2,embedding}. A Lie 2-algebra is an
  $L_\infty $-algebra concentrated only in degrees 0 and -1. If it is strict,
  it corresponds to a crossed module of Lie algebras \cite {Baez2}, which in
  turn corresponds to a crossed module of Lie groups. The present Lie 2-algebra
  is, however, non-strict and there is no evident relation to groups
  then.}\BibitemShut {Stop}%
\bibitem [{Note2()}]{Note2}%
  \BibitemOpen
  \bibinfo {note} {We propose the name 2-Higgs mechanism for this. First, it is
  a higher version of an ordinary Higgs mechanism, and in the language of
  `higher structures'' one conventionally labels a hierarchy of structures by
  an integer $n \in \protect \mathbb {N}$, with $n=1$ denoting the original
  notion. Second, and more importantly, it gives masses to 2-form gauge fields
  and, more generally, an $n$-Higgs mechanism would do so for $n$-form gauge
  fields.}\BibitemShut {Stop}%
\bibitem [{Note3()}]{Note3}%
  \BibitemOpen
  \bibinfo {note} {In fact, such an argument should also apply to the
  Leibniz-Chern-Simons theory considered in \cite {LeibnizCS}.}\BibitemShut
  {Stop}%
\bibitem [{Note4()}]{Note4}%
  \BibitemOpen
  \bibinfo {note} {As already evident from these examples at lowest levels, the
  truncation is not consistent, if one merely puts higher form-degree gauge
  fields to zero. Truncations always involve appropriate quotient constructions
  at the highest non-vanishing level. For example, the functional $S[A,0]$,
  resulting from \protect \textup {\hbox {\mathsurround \z@ \protect
  \normalfont (\ignorespaces \ref {S}\unskip \@@italiccorr )}} by setting the
  2-form fields to zero, is not gauge invariant anymore. But taking the
  quotient of $\protect \mathbb {V}$ by $\protect \mathbb {W}$, cf.\ \protect
  \textup {\hbox {\mathsurround \z@ \protect \normalfont (\ignorespaces \ref
  {sequence}\unskip \@@italiccorr )}}, we end up with the usual Yang-Mills
  functional at level one, $S_{YM}[A_\protect \mathfrak {g}]$.}\BibitemShut
  {Stop}%
\end{thebibliography}%


\begin{thebibliography}{99}
    \bibitem{talk}  \textsc{T.\ Strobl}, \emph{Mathematics around Lie 2-algebroids and the tensor hierarchy in gauged supergravity}. Talk at "Higher Lie theory", University of Luxembourg, 2013.
     \bibitem{Waldram} \textsc{K.\ Lee}, \textsc{C.\ Strickland-Constable} and \textsc{D.\ Waldram}, \emph{Spheres, generalized parallelisability and consistent truncations}, Fortsch.\ Phys.\ \textbf{65} (2017) 1700048.
      \bibitem{Lavau}\textsc{S.\ Lavau}, \emph{Tensor hierarchies and Lie n-extensions of Leibniz algebras}.  ArXiv:1708.07068, to be published in J.\ of Geom.\ and Physics.
      \bibitem{embedding} \textsc{A.\ Kotov} and \textsc{T.\ Strobl}, \emph{The Embedding Tensor, Leibniz-Loday Algebras, and Their Higher Gauge Theories}. ArXiv:1812.08611.   
\bibitem{H1}
 \textsc{H.~Nicolai} and  \textsc{H.~Samtleben},
\emph{Maximal gauged supergravity in three-dimensions},
Phys.\ Rev.\ Lett.\ \textbf{86} (2001) 1686.
\bibitem{H2}
 \textsc{B.~de Wit},  \textsc{H.~Samtleben} and  \textsc{H.~Trigiante},
\emph{On Lagrangians and gaugings of maximal supergravities},
Nucl.\ Phys.\ \textbf{B 655} (2003) 93.
\bibitem{H3}
\textsc{B.~de~Wit}, \textsc{H.~Samtleben} and \textsc{M.~Trigiante}, \emph{The
  maximal ${D} = 5$ supergravities}, Nucl.\ Phys.\ \textbf{B 716} (2005)
  215.
\bibitem{H4}
 \textsc{H.~Samtleben} and  \textsc{M.~Weidner},
\emph{The Maximal D=7 supergravities},
Nucl. Phys. \textbf{B 725} (2005) 383.
\bibitem{H5}
\textsc{B.~de~Wit} and \textsc{H.~Samtleben}, \emph{Gauged maximal
  supergravities and hierarchies of nonabelian vector-tensor systems},
  Fortschr. Phys., \textbf{53} (2005) 442.
\bibitem{H6}
\textsc{B.~de~Wit}, \textsc{H.~Nicolai} and \textsc{H.~Samtleben},
  \emph{Gauged supergravities, tensor hierarchies, and {M}-theory}, JHEP,
  \textbf{0802} (2008) 044.
 \bibitem{H7} \textsc{B.~de~Wit} and \textsc{H.~Samtleben}, \emph{The end of the $p$-form hierarchy}, JHEP \textbf{08} (2008)
  015.
  \bibitem{HenningohneLie4} \textsc{H.\ Samtleben}, \textsc{E.\ Sezgin} and \textsc{R.\ Wimmer},  
  \emph{(1,0) superconformal models in six dimensions}, JHEP \textbf{12} (2011) 062.
  \bibitem{Sam} \textsc{S.\ Palmer} and \textsc{C.~Saemann}, \emph{Six-Dimensional (1,0) Superconformal Models and Higher Gauge Theory}, J.~Math.~Phys.~\textbf{54} (2013) 113509.
\bibitem{Lie4} \textsc{S.\ Lavau}, \textsc{H.\ Samtleben} and  \textsc{T.\ Strobl},
\emph{Hidden Q-structure and Lie 3-algebra for non-\-abelian superconformal models in six dimensions},
  J.\ of Geom.\ and Physics {\bf 86} (2014) 497.
  \bibitem{gerbe}\textsc{T.\ Strobl},
\emph{Non-abelian Gerbes and Enhanced Leibniz Algebras.}
Phys.\ Rev.\  {\bf D 94} (2016) 021702.
\bibitem{Wagemann} \textsc{T.~Strobl} and \textsc{F.~Wagemann}, \emph{Enhanced Leibniz algebras: structure theorem and induced Lie 2-algebra}. ArXiv:1901.01014. 
 \bibitem{Baez} \textsc{J.\ Baez}, \emph{Higher Yang-Mills Theory}. ArXiv:0206130.
 \bibitem{Lada1}\textsc{T.~Lada} and \textsc{M.\ Markl},
\emph{Strongly homotopy Lie algebras}. 
Communications in Algebra 
{\bf 23} (1995) 2147.
\bibitem{Lada2}\textsc{T.~Lada} and \textsc{J.\ Stasheff}, 
\emph{Introduction to SH Lie algebras for physicists}. 
Int.\ J.\ Theor.\ Phys.\ \textbf{32} (1993) 1087.
\bibitem{Gruetzmann} \textsc{M.\ Gr\"utzmann} and \textsc{T.\ Strobl},
\emph{General Yang-Mills type gauge theories for p-form gauge fields: 
from physics-based ideas to a mathematical framework or from Bianchi 
identities to twisted Courant algebroids.} 
Int.\ J.\ Geom.\ Methods Mod.\ Phys.\ 
\textbf{12} (2015) 1550009.
\bibitem{Qbundles}  \textsc{A.\ Kotov} and \textsc{T.\ Strobl}, \emph{Characteristic classes associated to Q-bundles}.
      International Journal of Geometric Methods in Modern Physics {\bf 12}  (2015) 1550006. 
  \bibitem{Loday1} \textsc{J-L.\ Loday},
\emph{Une version non commutative des alg\`ebres de Lie: les alg\`ebres de
Leibniz.}
Enseign.\ Math.\ {\bf 39} (1993) 269.
\bibitem{Loday2} \textsc{J-L.\ Loday} and  \textsc{T.~Pirashvili}, 
\emph{Universal enveloping algebra of Leibniz algebras and (co)homology}.
Math. Ann. {\bf 296} (1993) 139.
\bibitem{W1} \textsc{S.\ Covez}, \emph{The local integration of Leibniz algebras}. Ann. Inst. Fourier (Grenoble) \textbf{63} (2013) 1.
\bibitem{W2}\textsc{J.\ Mostovoy}, \textsc{J.M.\ Perez-Izquierdo} and \textsc{I.P.\ Shestakov}, 
\emph{Hopf algebras in non-associative Lie theory}. 
Bull.\ Math.\ Sci.\ \textbf{4} (2014)129.
\bibitem{W3}\textsc{M.\ Bordemann} and \textsc{F.\ Wagemann}, 
\emph{Global integration of Leibniz algebras}. 
J. Lie Theory \textbf{27} (2017) 555. 
\bibitem{Lie21}\textsc{E.\ Getzler}, \emph{Lie theory for nilpotent $L_\infty$-algebras}. Ann. of Math. {\bf 170} (2009) 271.
\bibitem{Lie22}
\textsc{A.\ Henriques}, 
\emph{Integrating $L_\infty$-algebras}. 
Compos. Math. {\bf 144} (2008) 1017. 
\bibitem{Lie23} \textsc{P.\ Severa} and \textsc{M.\ Siran}, \emph{Integration of differential graded manifolds}. ArXiv:1506.04898.
\bibitem{LeibnizLie2} \textsc{Z.\ Liu} and \textsc{Y.\ Sheng},
\emph{From Leibniz algebras to Lie 2-algebras.}
Algebr.\ Represent.\ Theory {\bf 19} (2016) 1.
\bibitem{Baez2} \textsc{J.\ Baez} and \textsc{ A.\ Crans}, \emph{Higher-dimensional algebra. VI. Lie 2-algebras}. 
Theory Appl. Categ. \textbf{12} (2004), 492.
    \bibitem{LeibnizCS}\textsc{O.\ Hohm} and 
\textsc{H.\ Samtleben}, 
\emph{Leibniz-Chern-Simons Theory
and Phases of Exceptional Field Theory}. ArXiv:1805.03220. (To appear in Comm.\ Math.\ Phys.)
\end{thebibliography}
\end{document}